%% file: an_interesting_ODE.tex
\documentclass[aps,
prl,
english,
longbibliography,
showkeys,
reprint
]{revtex4-2}

\pdfoutput=1
\usepackage{amsfonts,amsmath,amssymb}
\usepackage{graphicx}
\usepackage[utf8]{inputenc}
\usepackage{hyperref}
\usepackage{babel}
\usepackage{enumitem}

\newcommand\be{\begin{equation}}
\newcommand\ee{\end{equation}}
\newcommand\bea{\begin{eqnarray}}
\newcommand\eea{\end{eqnarray}}

\begin{document}

\def\rhoo{\rho_{_0}\!} 
\def\rhooo{\rho_{_{0,0}}\!} 

\begin{flushright}
\phantom{
{\tt arXiv:2006.$\_\_\_\_$}
}
\end{flushright}


\title{
A differential equation for a class of correlation kernels}
\author{Clifford V. Johnson}
\email{cliffordjohnson@ucsb.edu}

\affiliation{Department of Physics, Broida Hall,   University of California, 
Santa Barbara, CA 93106, U.S.A.}


\begin{abstract}
A  new  differential equation is derived for an object ${\widehat S}(E,E^\prime,x)$, which when integrated over the appropriate range in $x$, yields the kernel $K(E,E^\prime)$ with which $n$--point correlation functions can be computed  in a wide class of models. 
When $E{=}E^\prime$, the equation reduces to the equation for the diagonal resolvent ${\widehat R}(E,x)$ of the Schr\"odinger Hamiltonian  ${\cal H}{=}{-}\hbar^2\partial_x^2{+}u(x)$ that is familiar from the classic work of Gel'fand and Dikii, and which appears in many areas of physics. This more general equation may also prove to be useful in a wide range of applications. Some special cases relevant to random matrix theory are explored using analytical and numerical methods.

\end{abstract}


\maketitle



{\it Introduction}---The problem of computing correlation functions 
in a wide range of physical problems sometimes boils down to finding a ``correlation kernel'', in terms of which the results can be expressed. Applications are to be found in the fields of probability and statistics, condensed and soft matter, number theory, geometry and topology,  random matrix models of nuclear physics, quantum gravity, quantum chaos, growth models, and many besides~\cite{Meh2004,ForresterBook,johansson2005}. These are often described as determinantal processes. For example, in a large class of random matrix models, in certain scaling limits, the eigenvalues $\{ E_i\}$ are distributed densely on the line, with density (one-point function) $\rho(E)$, given by the $E^\prime{=}E$ diagonal  of a kernel $K(E,E^\prime)$. The $n$--point function $\rho^{(n)}(\{E_1,\cdots,E_n\})$ is given by ${\rm det}||K(E_i,E_j)||_{i,j=1}^{n}$, where $||\cdots||$ denotes the $n{\times}n$ matrix. 
 Moreover, $K(E,E^\prime)$ appears as the kernel of an integral operator whose Fredholm determinant computes gap probabilities~\cite{Gaudin1961SurLL,Dyson:1962es}, an important tool that connects  probability theory, geometry, and ordinary differential equations of Painlev\'e type~\cite{Tracy:1992rf,*Tracy:1993xj}.

In these problems there is often  an organizing  associated Schr\"odinger problem on ${\mathbb R}$, with  Hamiltonian:
\begin{equation}
    {\cal H}=-\hbar^2\frac{\partial^2}{\partial x^2}+u(x)\ ,
\end{equation} where $u(x)$ is a potential  specific to the problem. The spectral density can be written in terms of the 
diagonal resolvent ${\widehat R}(E,x){=}\langle x|({\cal H}-E)^{-1}|x\rangle$ of $\cal H$: 
\begin{equation}
\label{eq:density-relation}
\rho(E)=K(E,E) =-\frac{1}{\pi\hbar}{\rm Im}\int_{a}^b \!{\widehat R}(E,x) dx\ ,
\end{equation}
(the integration range $(a,b)$ depends upon  the  problem), 
and ${\widehat R}(E,x)$ satisfies the  differential equation:
\begin{equation}
\label{eq:GD}
    4(u(x)-E){\widehat R}^2-2\hbar^2 {\widehat R}{\widehat R}^{\prime\prime}+\hbar^2({\widehat R}^\prime)^2 = 1\ ,
\end{equation}
familiar in a wide range of physics 
literature,  sometimes called the Gel'fand-Dikii equation~\cite{Gelfand:1976B,*Gelfand:1976A}\footnote{The equation played a key role in their foundational work on Sturm-Liouville theory, KdV integrable systems, and inverse scattering. In the form  given in Eq.~(\ref{eq:GD2}) it also appeared in ref.~\cite{marvcenko1974periodic} and (for $E{=}0$) in {\it e.g.} ref.\cite{LazutkinV.F.1976Nfav} around the same time.}. 

An important example is the case $u(x){=}{-}x$, for which the $x$-integration  is from ${-}\infty$ to $0$. In perturbation theory, the  leading solution to~(\ref{eq:GD}) is ${\widehat R}_0(E,x){=}{\pm} \frac12\sqrt{-x{-}E}$, which integrates  to give the leading spectral density $\rho_0(E){=}(\pi\hbar)^{-1}\sqrt{E}$, the scaled ``soft'' edge of the celebrated Wigner semi-circle law for Gaussian random Hermitian $N{\times} N$ matrices, where $N$ is large. Here,~$\hbar$ is the  $1/N$ expansion parameter. 
The asymptotic series of~$\hbar$ corrections to $\rho_0(E)$ is readily obtained by recursively expanding~(\ref{eq:GD}) about ${\widehat R}_0(E,x)$ and using~(\ref{eq:density-relation}), and it contains all the non-perturbative information as well.  This is often called the Airy model, since here the Schr\"odinger equation is simply Airy's equation~\cite{Moore:1990cn,Bowick:1991ky}.

For general $u(x)$, the  full kernel is built from the wavefunctions $\psi(E,x)$ of ${\cal H}$ (appropriately normalized), if they can be found:
\begin{equation}
\label{eq:introducing-the-kernel}
    K(E,E^\prime)=\int_{-\infty}^0\psi(E,x)\psi(E^\prime,x) dx\ .
\end{equation}
As seen above for the diagonal case, giving $\rho(E)$, the exact wavefunctions need not be known. The differential equation~(\ref{eq:GD})  supplies all that is needed to construct $K(E,E)$ perturbatively and non-perturbatively. 

It would therefore be extremely valuable to have an equation analogous to~(\ref{eq:GD}) that defines a more general function, denoted  ${\widehat S}(E,E^\prime,x)$, such that, extending~(\ref{eq:density-relation}): \begin{eqnarray}
\label{eq:getting-the-kernel}
    K(E,E^\prime) =-\frac{1}{\pi\hbar}{\rm Im}
    \int_{a}^b\!{\widehat S}(E,E^\prime,x) dx\ .
\end{eqnarray}
Having such an equation would allow for perturbative expansions of correlators to be systematically developed, as well as full non-perturbative information to be extracted.

This Letter proposes the following new 
differential equation defining the desired function ${\widehat S}(E,E^\prime,x)$:
\begin{eqnarray}
\label{eq:beyondGD}
      &&4\left(u-\frac{E+E^\prime}{2}\right){\widehat S}{\widehat S}^\prime + 4u^\prime {\widehat S}^2-2\hbar^2{\widehat S}{\widehat S}^{\prime\prime\prime}\\
      &&\hskip2cm+2(u-E){\widehat R}_1{\widehat R}_2^\prime+2(u-E^\prime){\widehat R}_2{\widehat R}_1^\prime= 0\ .
\nonumber\end{eqnarray}
 Here ${\widehat R}_{1}{\equiv}{\widehat R}(E,x)$ and ${\widehat R}_{2}{\equiv}{\widehat R}(E^\prime,x)$, which separately satisfy~(\ref{eq:GD}). When $E^\prime{=}E$, the new equation collapses back into~(\ref{eq:GD}), as can be seen by working with an equivalent form of~(\ref{eq:GD}) obtained by taking a derivative and removing a overall factor of~${\widehat R}$: 
\begin{equation}
\label{eq:GD2}
    4(u(x)-E){\widehat R}^\prime + 2u^\prime {\widehat R}-\hbar^2{\widehat R}^{\prime\prime\prime}= 0\ ,
\end{equation}
which is to be supplemented by the requirement that the  leading classical solution (obtained by setting $\hbar{=}0$) is ${\widehat R}_0(E,x){=}{\pm} \frac12\sqrt{u_0(x){-}E}\ ,$ fixing the integration constant on the right hand side in~(\ref{eq:GD}) to be unity.  (In some cases, $u(x)$ itself has an $\hbar$ dependence, and so $u_0(x)$  denotes its leading ($\hbar{\to}0$) piece of $u(x)$.) 
Now it can be seen that when the ${\widehat R}$s are both equal to ${\widehat S}$, the second line of (\ref{eq:beyondGD}) supplies an extra copy of the first term in the first line, and the system collapses back into  (\ref{eq:GD2}). 

 \medskip
 {\it Derivation of the Equation}---The derivation of the proposed form for the  central equation~(\ref{eq:beyondGD}) is simple, generalizing the procedure one can follow for deriving the form of~(\ref{eq:GD2}). In that case, one asks what equation the product $R{\equiv}\psi(x,E)^2$  solves. First, note that $R^{\prime\prime\prime}{=}6\psi^\prime\psi^{\prime\prime}{+}2\psi\psi^{\prime\prime\prime}$. On the other hand, the Schr\"odinger equation gives $\hbar^2\psi^{\prime\prime}{=}(u{-}E)\psi$, and $\hbar^2\psi^{\prime\prime\prime}{=}(u{-}E)\psi^\prime{+}u^\prime\psi\ .$ These can be used to eliminate the second and third derivatives of~$\psi$ in the above. Finally, $R^\prime{=}2\psi\psi^\prime$, which can be used to remove all appearances of $\psi$, resulting in~(\ref{eq:GD2}) for~$R$.

 Starting again with $S{\equiv}\psi(E,x)\psi(E^\prime,x){=}\psi_1\psi_2$,  following similar steps yields the intermediate stage:
 \begin{eqnarray}
     &&\hbar^2S^{\prime\prime\prime}=2u^\prime S+(u-E)[\psi_1^\prime\psi_2+3\psi_1\psi_2^\prime]\\
     &&\hskip1.8cm+(u-E^\prime)[\psi_1\psi_2^\prime+3\psi_1^\prime\psi_2]
    \nonumber\\
&&\hskip0.95cm    =2u^\prime S+
[2u-(E+E^\prime)] S^\prime
\nonumber\\&&\hskip1.8cm+2(u-E)[\psi_1\psi_2^\prime]+2(u-E^\prime)[\psi_2\psi_1^\prime]
    \ , \nonumber
 \end{eqnarray}
 where the second rewriting used $S^\prime {=} \psi_1^\prime\psi_2+\psi_1\psi_2^\prime$. To eliminate the occurrences of $\psi_{1(2)}$ it is enough to multiply by an overall $2S{=}2\psi_1\psi_2$, whereupon it is recognized that $2\psi_1^2\psi_2\psi_2^\prime\equiv R_1R_2^\prime$ and $2\psi_2^2\psi_1\psi_1^\prime\equiv R_2R_1^\prime$. Hence,~$S$ satisfies the equation of the proposed form for the full object, ${\widehat S}(E,E^\prime,x)$ given in~(\ref{eq:beyondGD}).

 \medskip
 {\it The Simplest Model}---The special  ``Bessel'' models have potential $u(x){=}{\hbar^2(\Gamma^2-\frac14)}/{x^2}$, where $\Gamma$ is an integer or half integer~\footnote{After a change of variables, it can be seen~\cite{Carlisle:2005wa} that the Schr{\" o}dinger problem becomes Bessel's equation, and the wavefunctions are made from Bessel functions of order $\Gamma$. These models appear naturally in random matrix models of Wishart form which have a ``hard'' edge~\cite{doi:10.1063/1.530157,FORRESTER1993709,Tracy:1993xj}. They have recently featured prominently in supersymmetric models of low dimensional gravity, and black hole dynamics.~\cite{Stanford:2019vob,*Johnson:2020heh,*Johnson:2020exp}\cite{Turiaci:2023jfa,*Johnson:2023ofr,*Johnson:2024tgg}}. If  $\Gamma{=}{\pm}\frac12$,  $u(x){=}0$  and the wavefunctions are simply trigonometric functions. Choosing a normalization, they are $\psi(E,x){=}A\sin(\frac{\sqrt{E}}{\hbar}x)$ for $\Gamma{=}\frac12$, and $\psi(E,x){=}A\cos(\frac{\sqrt{E}}{\hbar}x)$ for $\Gamma{=}{-}\frac12$, with $A^{-1}{=}{\sqrt{\pi\hbar}E^\frac14}$.
The integration over $x$ is from 0 to 1 in these models, and  the exact spectral density is:
\begin{equation}
\label{eq:bessel-density}
    \rho(E)=\int_0^1 \psi(E,x)^2 dx = \frac{1}{2\pi\hbar\sqrt{E}}\pm\frac{\sin(\frac{2\sqrt{E}}{\hbar})}{4\pi{E}}\ .
\end{equation}
In the small $\hbar$ expansion, there is a  leading perturbative piece that diverges at $E{=}0$,  all higher order terms   vanish, and there is  an oscillatory non-perturbative piece, which for $\Gamma{=}\frac12$ exactly cancels the divergence. The leading part of~(\ref{eq:GD2}) shows that ${\widehat R}$ starts out as $-\frac12(-E)^\frac12$, which is what yields the leading term in~(\ref{eq:bessel-density}), following from~(\ref{eq:density-relation}). In fact,~(\ref{eq:GD2}) is readily  solved using trigonometric functions in this case, since it is ${\widehat R}^{\prime\prime\prime}{=}{-}4E{\widehat R}^\prime/\hbar^2$. For $\Gamma{=}\frac12$:
\begin{eqnarray}
\label{eq:bessel-resolvent}
\nonumber
    {\widehat R}(E,x)&=&
    -\frac{1}{2\sqrt{-E}}\left(1-\cos({2\sqrt{E}\,x}/{\hbar})\right) \\
    &=& i{E}^{-\frac12}\sin^2({\sqrt{E}\,x}/{\hbar})\ .
\end{eqnarray} It is natural to place two copies of this exact solution (one for $E$ and one for $E^\prime$) into Eq.~(\ref{eq:beyondGD}) (with $u{=}u^\prime{=}0$) and seek a solution for ${\widehat S}$. After some algebra, an exact solution can be  found (for $\Gamma{=}{-}\frac12$ exchange sine for cosine):
\begin{eqnarray}
\label{eq:non-diagonal-exact}
    {\widehat S}(E,E^\prime,x) &=& -\frac{1}{(-E)^\frac14(-E^\prime)^\frac14}\times\\
    &&\hskip1cm\sin(\sqrt{E} x/\hbar)\sin(\sqrt{E^\prime} x/\hbar)\ , \nonumber
\end{eqnarray}
which  reduces to ${\widehat R}(E,x)$ in Eq.~(\ref{eq:bessel-resolvent}) when $E{=}E^\prime{=}E$. In retrospect, however, this had to be a solution given the derivation of the previous section. In other words, ${\widehat S}(E,E^\prime,x)$ is simply $ i\pi\hbar\psi(E,x)\psi(E^\prime,x)$ and there is no real part. Nevertheless, it serves as a useful warm-up case for a numerical treatment of the equation, treating it as a third-order non-linear system with exactly known ${\hat R}_{1(2)}$ inserted. Indeed, (for a range of sample energies) solution~(\ref{eq:non-diagonal-exact}) was obtained numerically to~(\ref{eq:beyondGD}) to good accuracy. More numerical work will be explored below. 
Using~(\ref{eq:getting-the-kernel}) the complete kernel for $\Gamma{=}{\pm}\frac12$  is:
\begin{eqnarray}
&&K(E,E^\prime) = \frac{(\sqrt{E}-\sqrt{E^\prime})\sin((\sqrt{E}+\sqrt{E^\prime})/\hbar)}{2\pi\hbar E^\frac14 {E^\prime}^\frac14(E-E^\prime)}\nonumber\\
&&\hskip1.0cm\mp\frac{(\sqrt{E}+\sqrt{E^\prime})\sin((\sqrt{E}-\sqrt{E^\prime})/\hbar)}{2\pi\hbar E^\frac14 {E^\prime}^\frac14(E-E^\prime)}\ ,
\end{eqnarray}
from which one can form the exact connected two point function {\it via} $\rho^{(2)}(E,E^\prime){=}{-}\hbar^2K(E,E^\prime)^2$. Incidentally, for large $E,E^\prime$ where it is natural to average over fast oscillations, this reduces to the universal formula~\cite{Brezin:1993qg,*BEENAKKER1,*BEENAKKER2,*BEENAKKER3,*Forrester1994}:
\begin{equation}
    \rho^{(2)}(E,E^\prime)\to -\frac{E+E^\prime}{4\pi^2\sqrt{E}\sqrt{E^\prime}(E-E^\prime)^2}\ ,
\end{equation}
to which $1/2\pi^2(E{-}E^\prime)^2$ can be added to restore good behaviour at $E{=}E^\prime$.

 \medskip
 {\it Leading Solutions}---The leading order solution, ${\widehat S}_0$, (say, at large $x$, or small $\hbar$) of  Eq.~(\ref{eq:beyondGD}) is obtained by dropping the $\hbar^2$ term, yielding a first order equation:
 \begin{equation}
 \label{eq:bernoulli}
      {\widehat S}_0^\prime = f^{-1}\left[{g}{\widehat S}_0^{-1}-4u_0^\prime {\widehat S}_0\right]\ ,
 \end{equation}
 where $g,f$ and  $u_0$ 
(the leading piece of $u(x)$) depend on~$x$:
 \begin{eqnarray}
     &&f(x) = 2\left(2u_0-(E+E^\prime)\right)\ ,\\
     &&g(x) = \frac{u_0^\prime}{4}\left[\frac{(u_0-E)^2+(u_0-E^\prime)^2}{(u_0-E)^\frac32(u_0-E^\prime)^\frac32}\right]\ ,
 \end{eqnarray}
and leading form ${\widehat R}_0(E,x){=}{-}\frac12(u_0-E)^{-\frac12}$ was used for each energy. (When $u(x){=}0$, the only solution is constant~${\widehat S}_0$, seen in the previous section.) Eq.~(\ref{eq:bernoulli}) is of   Bernoulli form, and so a first pass at the nature of solutions can be found by the linearizing  substitution  $T(x) {=} {\widehat S}_0(x)^2$, yielding the equation $T^\prime {=} f^{-1}(2g-8u_0^\prime T)$. Upon inspection the solution is
$T(x) {=} \frac{2}{f(x)^2} F(x)$, where
$F(x){\equiv}\int^x \!\!f(y)g(y)dy{+}c_2$, with $c_2$  a constant, and so: 
  \begin{equation}
  \label{eq:tee}
     {\widehat S}_0(E,E^\prime,x)=\pm\frac{\sqrt{2F(x)}}{f(x)}\ .
 \end{equation}
As a specific example, recall the  Airy model  (see below~(\ref{eq:GD})), where   $u(x){=}{-}x$.  The  wavefunctions are $\psi(E,x){=} \hbar^{-\frac23} {\rm Ai}(-\hbar^{-\frac23}(E+x))$, with normalization  that yields the leading spectral density $\rho_0(E){=}(\hbar\pi)^{-1}\sqrt{E}$. 
In this case, a calculation gives:
\begin{eqnarray}
 F(x)=  
    \frac{4(x^2+EE^\prime)-(x-E)^2-(x-E^\prime)^2}{(-x-E)^\frac12(-x-E^\prime)^\frac12} {+} c_2\quad
\end{eqnarray}
 and so (picking the right sign of the root of $T(x)$):
\begin{eqnarray}
\label{eq:asymptotic-form-for-big-equation}
{\widehat S}_0(E,E^\prime,x) = -\frac12\frac{1}{(-x-E)^\frac14(-x-E^\prime)^\frac14}\times G(x)\ ,\quad\,
\end{eqnarray}
where the function $G(x)$  is unity when $E^\prime{=}E$, and in that case, this result has the same form as one of the ${\widehat R}_0(E,x)$, as the full Eq.~(\ref{eq:beyondGD}) guarantees. (The constant $c_2$ can be taken to be zero here, or to also vanish when $E^\prime{=}E$.)
Moreover,  $G(x){\to}1$ at large $x$,
giving the same leading  form as  $\psi(E,x)\psi(E^\prime,x){=}{\rm Ai}(-(E{+}x)){\rm Ai}(-(E^\prime{+}x))$ (up to a normalization and before taking the imaginary part). This form of ${\widehat S}_0(x)$ (and other examples for more general~$u_0(x)$) can be used as asymptotic large~$x$ boundary conditions to the full differential equation~(\ref{eq:beyondGD}) in order  to construct fully non-perturbative solutions (See below).

An even more general solution of~(\ref{eq:bernoulli}) can  be found by considering complex solutions of equations of Bernoulli type. Recent work~\cite{axioms13010026} found  exact solutions by exploiting connections to Lie-Hamilton systems. The case in hand sets their $(t,n,a_1,a_2)$ to $(x,{-}1,{-}f^\prime/f,g/f)$, which   yields a very simple form.  Writing ${\widehat S}{=}
{\widehat S}_0^+{+}i{\widehat S}_0^-$ it is:
\begin{equation}
    {\widehat S}^\pm_0(E,E^\prime,x)=-\frac{1}{\sqrt{2c_1}f(x)}
\sqrt{(1+B^2)^\frac12\pm B}\ ,
\end{equation}
where  $B{\equiv} 2c_1F(x)$, $c_1$ is a constant, and $F(x)$ is defined above~(\ref{eq:tee}). At large $B$, ({\it e.g.,} for large $x$ or $E,E^\prime$), ${\widehat S^{-}_0}$ vanishes and ${\widehat S_0^+}$ becomes the result obtained in~(\ref{eq:tee}).

\begin{figure}
    \centering
    \includegraphics[width=0.95\linewidth]{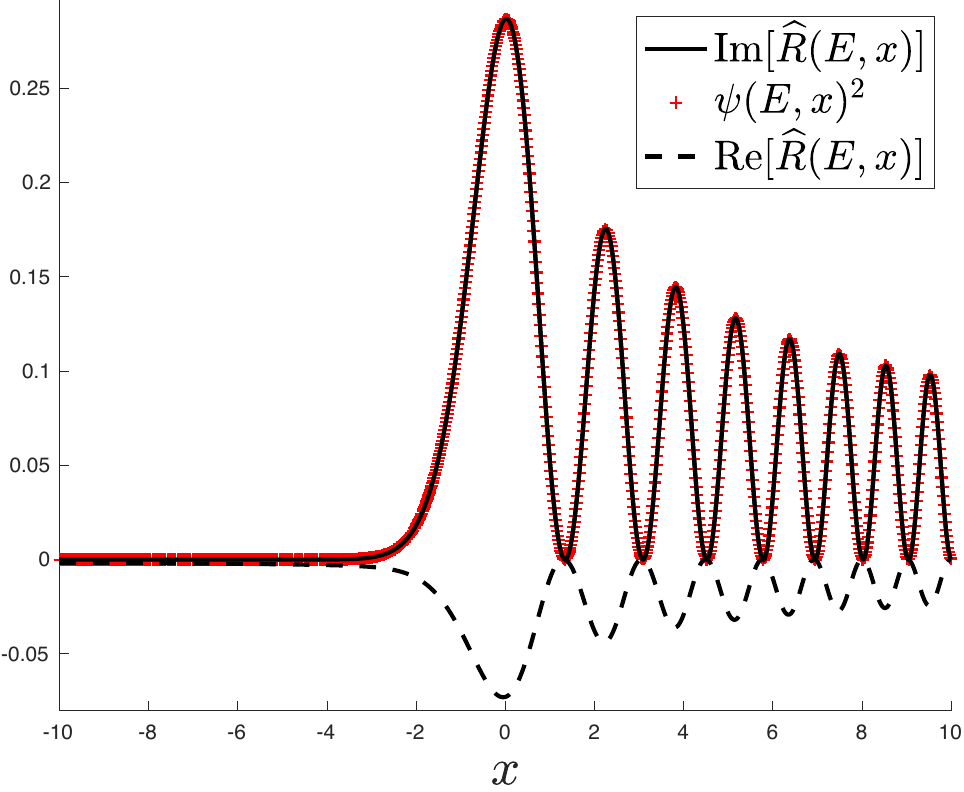}
    \caption{Imaginary part of the solution  to Eq.~(\ref{eq:GD2}) for ${\widehat R}(E,x)$, compared to $\psi(E,x)^2$, up to a normalization. The real part is also shown. Here~$E{=}1$, and $\hbar{=}1$.}
    \label{fig:GD-solution}
\end{figure}

\medskip
{\it Numerical Solutions}---Eq.~(\ref{eq:beyondGD}) is highly non-linear, making it a delicate matter to numerically handle the necessarily oscillatory solutions stably. It was  mentioned above that the simple purely imaginary exact solution for the $u(x){=}0$ Bessel case was used to test the methods. It is important to explore solutions that are not purely imaginary ({\it i.e.,} not a simple product of wavefunctions).  The Airy model $u(x){=}{-}x$ will be used. As a warm-up, it is worthwhile to reflect on the diagonal case again. The relevant differential equation~(\ref{eq:GD2}) can be provided with the large~$x$ boundary condition ${\widehat R}{=}{-}\frac12(-x{-}E)^\frac12$ for some positive~$E$. To the far left this is a good solution, while to the right it describes only the average fall off of an oscillatory behaviour, but it is still a good enough indication of the solution's size for large enough $x$. The equation is robust enough to not be too sensitive to the error in not inputting the oscillatory form (but see later). This is all easily set up as a boundary value problem in {\tt MATLAB}, and a typical result (solved in about $1.6$ seconds on a grid of ${\sim}8600$ points on $[-40,40]$) is shown (both real and imaginary parts) in Fig.~{\ref{fig:GD-solution}}. 
The black line is the imaginary part of the computed solution while the red crosses are the comparison to $\psi(E,x)^2$ (up to a normalization) and the results are of course in excellent agreement.

Emboldened by this success, turning to  the new Eq.~(\ref{eq:beyondGD}), for a pair of energies $(E, E^\prime)$,  note that it takes as input a coupling to two functions ${\widehat R}(E,x)$ and ${\widehat R}(E^\prime,x)$,  solutions of the simpler  Eq.~(\ref{eq:GD2}). Rather than solving  numerically for the ${\widehat R}$s separately and then using them as input, it is wiser (in view of potential numerical error for such an unstable system) to numerically solve the entire system as a set of three coupled third order differential equations (hence, a ninth order system) as a boundary value problem in {\tt MATLAB}~\footnote{Some {\tt MATLAB} details: The solver {\tt bvp4c} was used, with the system presented as a set of 9 coupled 1st order equations. Given  the increased complexity and the non-linearity, vectorization was used, and the Jacobians  were input analytically, dramatically increasing the speed and accuracy. Solving to tolerance ${\sim}10^{-6}$ on a fast laptop typically took from 1s-5m on grids of 2K-20K points, depending upon parameter values.}. The leading form~(\ref{eq:asymptotic-form-for-big-equation})  (with $G(x){=}1$) was used for the boundary condition. 

\begin{figure}[t]
    \centering
    \includegraphics[width=0.95\linewidth]{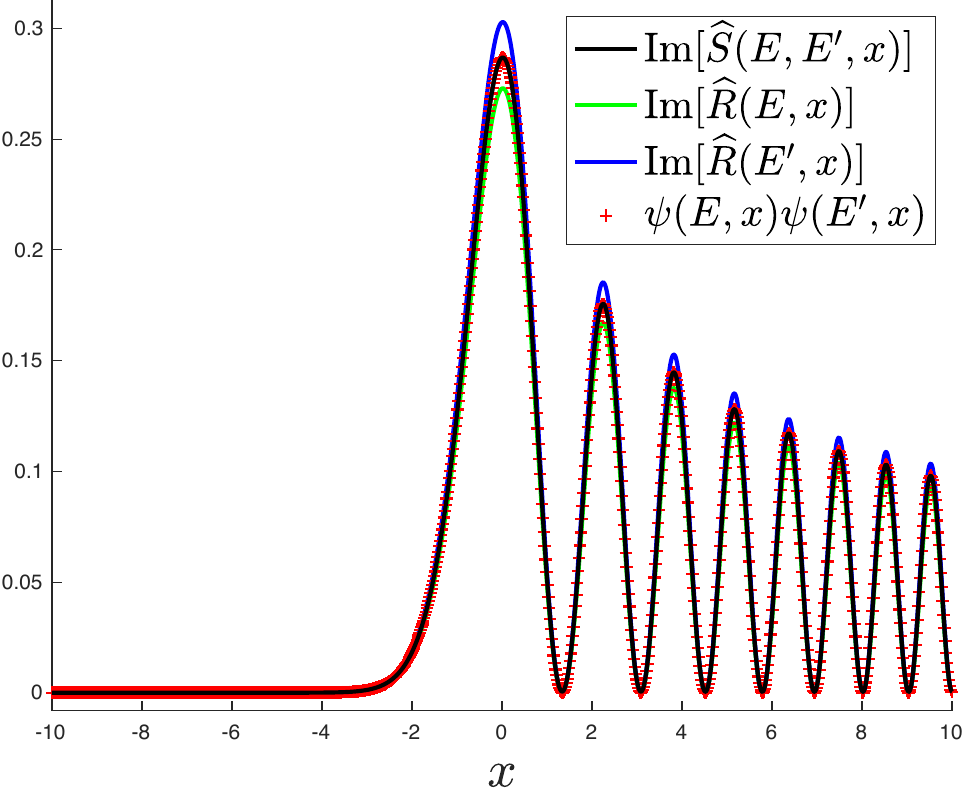}
    \caption{Imaginary part of the solution for ${\widehat S}(E,E^\prime,x)$ to Eq.~(\ref{eq:beyondGD}), compared to $\psi(E,x)\psi(E^\prime,x)$, up to a normalization. Here $E{=}1$, $E^\prime{=}1.001$, and $\hbar{=}1$. 
    }
    \label{fig:new_cvj-solution1}
\end{figure}

\begin{figure}[t]
    \centering
    \includegraphics[width=0.95\linewidth]{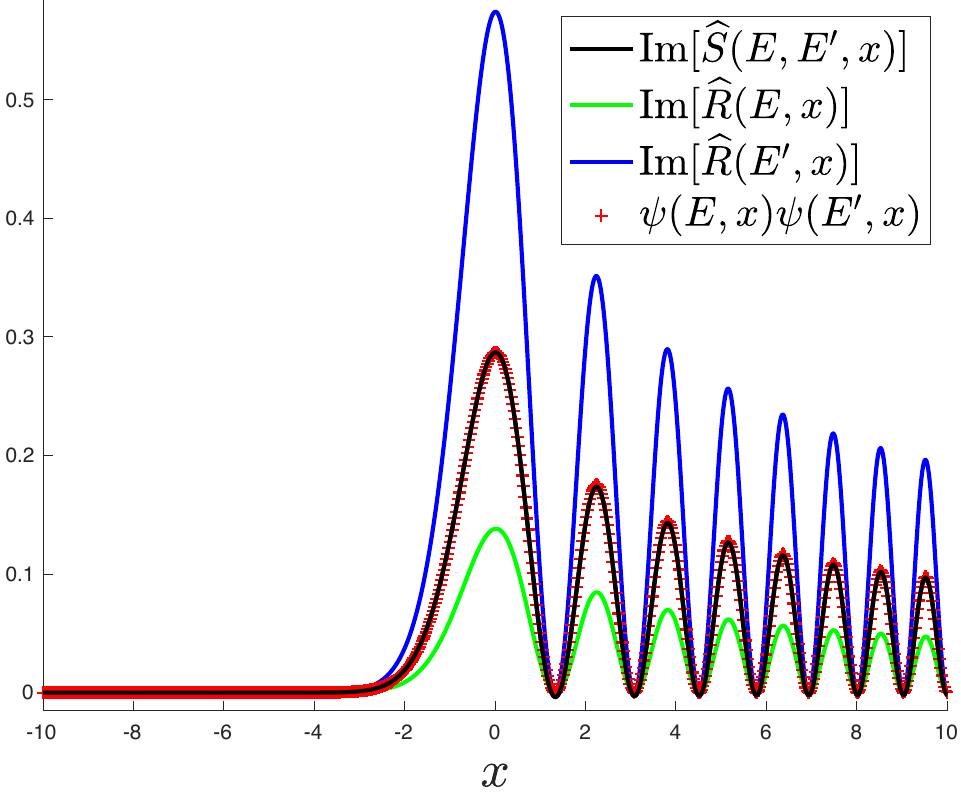}
    \caption{Imaginary part of the solution for ${\widehat S}(E,E^\prime,x)$ to Eq.~(\ref{eq:beyondGD}), compared to $\psi(E,x)\psi(E^\prime,x)$, up to a normalization. Here $E{=}1$, $E^\prime{=}1.01$, and $\hbar{=}1$. 
    }
    \label{fig:new_cvj-solution2}
\end{figure}

As a first test, the case of $E{=}E^\prime{=}1$ was solved to check that ${\widehat S}(E,E,x)$ coincides with the (now equal) ${\widehat R}(E)$s. Happily, the resulting figure for all functions is identical to Fig.~\ref{fig:GD-solution}. Introducing a 0.1\% difference, $E{=}1$, $E^\prime{=}1.001$, the resulting three functions can be seen to move apart,  as shown in Fig.~\ref{fig:new_cvj-solution1}. 
Fig.~\ref{fig:new_cvj-solution2} shows the result of increasing the difference in energies tenfold, with $E{=}1$, $E^\prime{=}1.01$. Strikingly, in all cases, a numerical check (shown) confirms that the imaginary part of ${\widehat S}(E,E^\prime,x)$ coincides with $\psi(E,x)\psi(E^\prime,x)$!
As the difference in energies increases, the problem becomes increasingly unstable, and going much beyond a 10\% difference becomes hard to control (although some examples can be done with techniques such as continuation), but the pattern is clear: The function ${\widehat S}(E,E^\prime,x)$ defined through the new differential equation~(\ref{eq:beyondGD}), gives a definition (through its imaginary part) of $\psi(E,x)\psi(E^\prime,x)$, as desired. 

As a further test, note that for general $\psi(E,x)$, the kernel~(\ref{eq:introducing-the-kernel}) can be written in Christoffel-Darboux  form as:
\begin{equation}
\label{eq:CD-form-of-kernel}
K(E,E^\prime)=\frac{\psi(E)\psi^\prime(E^\prime)-\psi^\prime(E)\psi(E^\prime)}{E-E^\prime}\ .
\end{equation}
For the  numerical examples above,  the  integral of $\pi^{-1}{\rm Im}[{\widehat S}(E,E^\prime,x)]$ from the left $x$-boundary to $x{=}0$ was compared directly to  this, with good agreement.

\medskip

{\it Closing Remarks}---The new differential equation~(\ref{eq:beyondGD}) defines an interesting function, ${\widehat S}(E,E^\prime,x)$, which naturally generalizes the familiar resolvent ${\widehat R}(E,x)$.  It is a non-diagonal departure from the resolvent in a different variable than is typically explored ({\it i.e.,} it is {\it not} the Green's function ${\widehat R}(E,x,x^\prime)$). As such, it is worth further exploring in its own right. It is  interesting that the leading solution, ${\widehat S}_0$ satisfies a complex Bernoulli equation of a special form, allowing a simple exact solution to be written with the aid of recent connections~\cite{axioms13010026} to Lie-Hamilton systems.  Perhaps there is more to be learned from those connections---Obtaining higher order $\hbar$ corrections to ${\widehat S}(E,E^\prime,x)$
involves solving additional first order differential equations. 

With Eq.~(\ref{eq:getting-the-kernel}), ${\widehat S}(E,E^\prime,x)$ gives an alternative definition of the correlation kernel $K(E,E^\prime)$, without the need for direct knowledge of the explicit wavefunctions. This could have applications in a wide range of systems (specified by  potential~$u(x)$) where $K(E,E^\prime)$ is useful.  There are many directions to pursue,  such as developing expressions for corrections (in an $\hbar$ expansion) to ${\widehat S}(E,E^\prime,x)$ beyond the leading order form uncovered here. Wider numerical exploration would also be useful. 

One motivation for finding  Eq.~(\ref{eq:beyondGD}) was the following. As already mentioned, for the (one-point) diagonal case, the Gel'fand-Dikii equation allows for a recursive development of perturbative (in $\hbar$) corrections to ${\widehat R}(E,x)$. It was recently observed~\cite{Johnson:2024bue,Johnson:2024fkm} that after integration they yield at order $\hbar^{2g-1}$ the quantities $W_{g,1}(z)$, ($z^2{=}{-}E$), which Laplace transform into ``volumes'' $V_{g,1}(b)$ associated to the moduli space of certain 2D geometries  with~1 boundary ($b$ is its length) and genus~$g$.  
For the  $u(x)$ (entered perturbatively) appropriate for hyperbolic Riemann surfaces~\footnote{In this case $u_0$ satisfies $\sqrt{u_0} I_1(2\pi\sqrt{u_0}){+}x{=}0$~\cite{Okuyama:2019xbv}, and perturbative corrections in $\hbar$ may be readily obtained by expanding the string equation satisfied by $u(x)$. See ref.~\cite{Johnson:2024bue} for how this is all used.},  $V_{g,1}$ are the Weil-Peterson volumes for the case of one boundary. Evidently, expanding  Eq.~(\ref{eq:GD}) is an efficient method of deriving them, equivalent to (but different from)  topological recursion~\cite{Mirzakhani:2006fta,Eynard:2007fi}\footnote{Ref.~\cite{Lowenstein:2024gvz} explores the connections somewhat.}.  With a new equation~(\ref{eq:beyondGD}) in hand that yields small~$\hbar$ perturbation theory for $K(E,E^\prime)$, for a given $u(x)$, it is in principle possible to develop expressions for the full $W_{g,n}(\{z_i\})$, (using an $n{\times}n$ determinant) and hence $V_{g,n}(\{b_i\})$, ($i{=}1,{\cdots},n$). The equation could thus allow powerful statements to be made about large genus behaviour of wide classes of volumes, and of course supply non-perturbative data too.

{\it Acknowledgments}---CVJ  thanks  the  US Department of Energy (\protect{DE-SC} 0011687) for  support,  and  Amelia for her support and patience.    









\input{an_interesting_ODE.bbl}

\end{document}

%% file: an_interesting_ODE.bbl
%